\documentclass[%
 reprint,
nofootinbib,
 amsmath,amssymb,
 aps,
]{revtex4-2}

\usepackage{graphicx}
\usepackage{dcolumn}
\usepackage{bm}
\usepackage{color}

\begin{document}


\title{Boundary-induced classical Generalized Gibbs Ensemble with angular momentum 
}

\author{Francesco Caravelli}
 \email{caravelli@lanl.gov}
\author{Marc D. Vuffray}%
 \email{vuffray@lanl.gov}
\affiliation{%
Theoretical Division (T4)\\Los Alamos National Laboratory,\\
Bikini Atoll road, Los Alamos 87505, NM USA
}%


\date{\today}

\begin{abstract}
We investigate the impact of the boundary shape on the thermalization behavior of a confined system of classical hard disks at low packing fraction and thus in the gas regime. We use both analytical calculations and numerical simulations, and leveraging on the insights from the maximum entropy principle, we explore how the geometry of the boundary influences the thermal equilibration process in such systems. Our simulations involve hard disks confined within varying boundary shapes, using both event-driven and time-driven simulations, ranging from conventional square boundaries to circular boundaries, showing that the two converge to different ensembles. The former converges to the Gibbs Ensemble, while the latter converges to the Generalized Gibbs Ensemble (GGE), with angular momentum as the extra conserved quantity. We introduce an order parameter to characterize the deviations from the Gibbs ensemble, and show that the GGE is not time-reversal invariant, it violates ergodicity and leads to a near-boundary condensation phenomenon.
Because of this, we argue that Monte Carlo methods should include angular momentum in this situation. We conclude by discussing how these results lead to peculiar violations of the Bohr-van Leeuwen theorem.


\end{abstract}

\maketitle


\section{Introduction}
Investigating how systems converge to the Gibbs distribution is a central focus in the study of many-body dynamics, spanning both classical and quantum domains. Statistical mechanics can be summarized as the ability to find an ensemble describing the statistical properties of physical observables as functions in phase space. Gibbs distributions emerge as a result of the fact that, given the assumptions of the ergodic hypothesis, energy is the only physical quantity that is conserved \cite{Kinchin}. However, this is not always the case. The goal of this manuscript is to show the importance of angular momentum in a simple model, and how it leads to a condensation phenomenon. In particular, we would like to highlight the importance of boundaries, which in the literature have had a much less relevant role in numerical simulations.

Over the last few decades, there has been significant interest in the behavior of quasi-integrable quantum systems that do not thermalize to a Gibbs ensemble. The Generalized Gibbs Ensemble (GGE) is employed when the equilibrium distribution relies on various conserved quantities beyond energy. Such phenomena may be more prevalent in quantum systems  \cite{rigol} compared to classical ones (except for some recent studies \cite{Casiulis,barbier}), as supported by recent studies. Notably, the Quantum Generalized Gibbs Ensemble has been experimentally observed \cite{langen}, highlighting its relevance in quantum statistical mechanics. 
This manuscript shows that it is quite easy to observe a classical GGE in the simplest model in statistical mechanics, by a slight change in the boundary.
We consider a 2D particle model characterized by
1) free propagation between collisions, 2) perfect reflections with the walls of the box, and 3) elastic collisions between particles, e.g. momentum and energy are conserved. 
An example of this type is, as mentioned in \cite{reviewhd}, the Hard Disk Model (HDM), which can be considered as the Drosophila of statistical mechanics. In this model, the disks are impenetrable and can also be thought as the simplest model of 
granular matter. Granular matter consists of discrete solid particles, ranging in size from hundreds of micrometers to several kilometers, that interact primarily through volume exclusion. These particles are generally too large to be significantly affected by thermal fluctuations \cite{Duran2000,Jaeger1996,deGennes1999,Kakalios2005,krauthb}.

The HDM has been studied with tens of thousands of particles using Metropolis algorithms or more recently using Event Chain Monte Carlo \cite{krauthecmc0,krauthecmc}.  The phenomenology of this model is very rich. There is a critical density for hard disks, separating the fluid and hexatic phase, and $\eta\approx 0.716$ for the critical transition density between hexatic and solid phases \cite{pt00,pt0,pt1,pt2,pt3,pt4,pt5,pt6}. Our simulations will be conducted in the gas phase, at very low packing fractions, in order to highlight that our results are a product of the boundary shape. 

The HDM is one of the most well-known models in the study of thermalization in statistical mechanics \cite{pulvirenti}. This is also one of the oldest models studied in the history of statistical mechanics \cite{bernoulli}.
The model was initially studied numerically with earlier computers, in particular ENIAC and MANIAC, at the Los Alamos Scientific Laboratory \cite{lanl0,lanl1,lanl2,lanl3,lanl4,lanl5,lanl6}, the scientific laboratory that preceded the modern Los Alamos National Laboratory. The study of hard sphere models had focused, initially, on square enclosures with reflecting or toroidal boundary conditions. It is known that at high densities, the hard-sphere gas undergoes a phase transition from a liquid to a solid phase, with virial coefficients now calculated up to the 10th order \cite{reviewhd}. It is in fact known that at high packing fractions, the system organizes in crystals \cite{pack1,pack2,pack3}.

Historically, and as far as we understand, the role of the confining boundary in the study of the thermalization of many body systems was somewhat neglected because one needs a fine-tuned boundary for extra conservation laws to appear \cite{fowler,dubrovskyi}. In addition, it is easy to use periodic boundary conditions, which do not conserve angular momentum.  Although the conservation law of angular momentum has been investigated in the literature \cite{drubovskyi2}, it has not been studied in simple models like the HDM. This fact is only anecdotally mentioned in the literature. For instance, the conservation of angular momentum with a circular boundary was mentioned in \cite{freezing}, but then the angular momentum was set to zero. That case, near the freezing point, is a completely different and non-kinetic regime from our naive and simple gas. However, one might ask questions of what happens when the boundary is restricted over time when angular momentum is conserved, and for instance, ask what happens at the freezing point in the presence of angular momentum.

Near freezing, traditional statistical mechanics methods are not applicable to granular matter, which is either static or out of equilibrium. Both states exhibit unique phenomena. In the static state, the macroscopic size and disordered packing of particles lead to inhomogeneous stress distributions and the formation of arches that direct loads sideways \cite{Zuriguel2007}. This results in pressure saturation inside containers (Janssen's law) and spontaneous jamming \cite{Janssen1895} or clogging in the presence of obstacles \cite{reichhardt}. In the dynamic state, the key effects are related to the system being out of equilibrium \cite{Liu1998}. These regimes require careful simulations \cite{perez}, and as far as we understand, there are disagreements on what it entails to simulate a hard disk model. Since in this case the devil is in the detail, it makes sense to study the much simpler low-packing fraction case for which basic methods will work and statistical mechanics can be applied, to show that even in this regime the asymptotic distributions are far from Gibbs.

An important point to make is that particle-particle interactions are fundamental in what we discuss. If particles do not interact, then with a circular boundary we simply have a collection of integrable motions. In this case, the total energy and angular distribution is of the form of a Dirac comb in the energy and angular momentum summed over each particle. In the context of circular billiards, the integrability of a single trajectory is attributed to the conservation of the angle of reflection at each boundary collision \cite{circ1,kalsorr}, which in turn is attributable to the conserved energy and angular momentum of the single particle. This constant angle serves as an integral of motion, leading to a foliation of the phase space into invariant curves. Each trajectory corresponds to a caustic, concentric circle within the billiard table, illustrating that the system can be described exactly through these conserved quantities. Consequently, the motion is predictable and the system is integrable due to these geometric and dynamical properties. For this reason, simulations with interactions are fundamental to achieving ergodicity. In our simulations, interactions will be of the simplest type, e.g. through excluded volumes. Particles cannot be pointlike, and they need to possess a finite volume.

For the reasons above, in two dimensions, it makes sense to focus on hard disk models. In different situations, Time-Driven Molecular Dynamics~(TDMD) and Event-Driven Molecular Dynamics~(EDMD) simulations have different pros and cons in more advanced simulations for high-density regimes.
These regimes, although interesting, are not the focus of this manuscript, where we focus on low packing fractions, and where there is an agreement on what are true simulations of hard disks \cite{Haile,Allen,Frenkel,algs4}, but we will still use both EDMD and TDMD to show the importance of the boundary on the asymptotic 1-body particle distributions with two different methods, in the absence of any frictional force and perfect slip. 

The goal of this manuscript is to show that boundaries play an important role in thermalization.  We study the asymptotic 1-body particle distribution both analytically and numerically, in the regime of low-density particles. Specifically for the HDM, if the boundary is circular and particles are reflected perfectly with respect to the tangent at the point of collision, the asymptotic distribution ceases to be Gibbs, as angular momentum is conserved. This implies a hard breakdown of ergodicity, and the system's asymptotic behavior becomes dependent on the initial condition.  In addition, the equilibrium distributions are not time-reversal invariant because of the presence of angular momentum.

\begin{figure}
\includegraphics[width=0.99\linewidth]{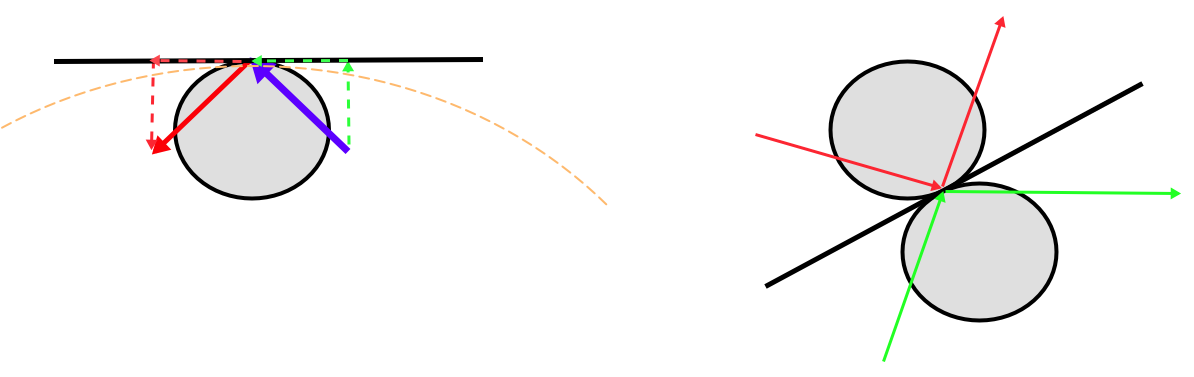}
\caption{The two types of collisions that we considered numerically. The first is a collision with the boundary, which reflects the component perpendicular to the tangent plane evaluated at the point of impact. The second is the elastic collision between the particles. In this collision, both momentum and energy are conserved, and there are no frictional dissipative forces between the particles. Also, in the collision, no torque is applied and thus particles that start with no angular momentum, remain at zero angular momentum, e.g. there is no rotation after a collision.}
\label{fig:collisions}
\end{figure}

This manuscript is organized as follows. We first introduce an order parameter to understand the results as a function of this order parameter. We then provide an interpretation of the results via the MaxEnt principle and show that the results are consistent with angular momentum conservation and the Generalized Gibbs Ensemble with two Lagrange multipliers, instead of a single one with the energy as the sole conserved quantity.  We provide numerical evidence that supports the correctness of our predictions. In the main text, we have used a naive EDMD simulation to test that the 1-body distributions do not converge to Gibbs in the long-time regime. In the appendix, we also report TDMD simulations and study the effect of eccentricity on the decay of the order parameter. We then apply this asymptotic distribution to show how to change a simple Monte Carlo algorithm to obtain the correct asymptotic distribution. We also discuss why the conservation of angular momentum would result in a violation of the Bohr-van Leuween theorem. Conclusions follow.

\begin{figure*}
    \centering
    \includegraphics[width=0.99\linewidth]{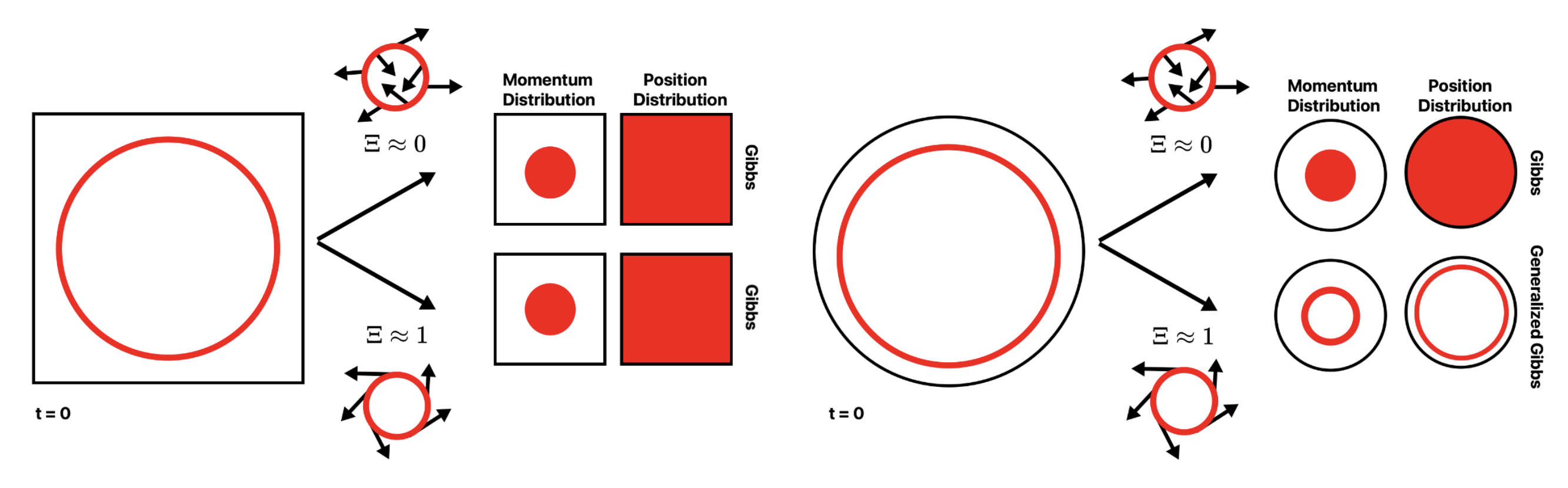}
    \caption{A summary of the results of this paper, categorized according to the order parameter $\Xi$ introduced in eqn.~\eqref{eq:opxi}. We consider the cartoon of the initial conditions, e.g. particle positions initialized on a thin disk, (red circular lines) both for the square and circular boundary (left and right respectively), but with the angular directions completely random in $[0,2\pi]$, and velocity set to a particular value $v_{in}$, leading to effectively zero angular momentum, $\Xi\approx 0$. If the angular directions are instead tangential to the circle, and oriented clockwise, then there is non-zero angular momentum, $\Xi>0$. For $\Xi=0$, the 1-body momentum and position distributions follow Gibbs. For $\Xi\approx 1$, the deviations from Gibbs are significant, the 1-body momentum and position distribution do not factorize, and the system experiences a condensation-near-the-boundary phenomenon of eqn.~\eqref{eq:condensation}.}
    \label{fig:summary}
\end{figure*}
\section{Statistical mechanics of particle gas with conserved angular momentum }

In this section, we focus on the statistical mechanics of a free gas in the presence of angular momentum conservation. We first introduce a conserved order parameter $\Xi$ that is a combination of the two conserved quantities assumed in this paper, energy and angular momentum. In the next sections, we show how this parameter is connected to both the MaxEnt probability and Canonical ensemble, and how it emerges from the microcanonical ensemble. We show that a condensation phenomenon emerges in the limit $\Xi\rightarrow 1$.

\subsection{Order parameter for particle systems with conserved energy and angular momentum }

The particle model we consider is rather simple to understand and is one of the first models considered in statistical mechanics courses and molecular dynamics \cite{Haile}. 
We keep the discussion generic for the time being, and we will go into the details of a model later. We consider a system of particles whose total energy is defined as $E$. The system is contained in an enclosure of size $R$. In the case in which the enclosure is a square, $R$ represents the size of each side of the square, while it represents the radius for a circular boundary. The system can have total angular momentum $L$ as the initial condition, calculated for simplicity with respect to the geometrical center of the enclosure. Each particle's coordinate is defined as $\mathbf{x}_i=(x_i,y_i)$, while the velocity vector as $\mathbf{v}_i=(v_i^x,v_i^y)$. We will use the simplified notation for the vector 2-norm $\|\mathbf{x}\|^2=x^2+y^2$. We will use the notation $r_i=\sqrt{\|\mathbf{x}_i\|^2}$, $p_i=\sqrt{\|\mathbf{p}_i\|^2}$ and $v_i=\sqrt{\|\mathbf{v}_i\|^2}$ going forward.

The key interactions we consider are of the type shown in Fig. \ref{fig:collisions}. These are either collisions between particles and a boundary, or collisions between particles. Our assumptions is that: a) Energy is conserved b) Momentum is conserved in particle collisions c) angular momentum is conserved in perfect reflections from the boundary. In particular, in the simulations we will do later, the particles we consider do not slip at collisions and there are no dissipative or active forces. In principle, there could be a central potential with which the particles are interacting, but this potential must be rotationally invariant. If however, this potential is not present, the only contribution to the energy is kinetic. In this case, we argue that the asymptotic behavior of the system is determined according to the following order parameter: 

\begin{eqnarray}
    \Xi=\frac{\|\sum_{i=1}^N {\bm x}_i \wedge {\bm p}_i\|^2}{2 R^2 (\sum_{i=1}^N m_i) (\sum_{j=1}^N \frac{\|\mathbf{p}_j\|^2}{2 m_j})}.\label{eq:opxi}
\end{eqnarray}
First, we will prove that in the case of conserved angular momentum:
\begin{eqnarray}
\frac{d}{dt} \Xi&=&0\\
\text{and}\ \ \ &&    0\leq \Xi\leq 1
\end{eqnarray}
In particular, $\Xi=0$ if angular momentum is zero, and $\Xi=1$ if the angular momentum is maximized for a given total kinetic energy. As we show later, this order parameter controls the deviation from the Gibbs ensemble.  A summary of how our results depend on $\Xi$ is sketched in Fig.~\ref{fig:summary}.

Let us now show why the order parameter above characterizes the state of the system. It is easy to see that since it is composed of only energy and angular momentum, it is a constant of motion and positive when these are conserved. To see that $|\Xi|\leq 1$, note that
\begin{eqnarray}
    L^2&=&\|\sum_{i} m_i {\bm x}_i\wedge {\bm v}_i\|^2\\
    &=&\left(\sum_{i}  \sqrt{m_i}\frac{{\bm x}_i\wedge {\bm v}_i}{\|\bm x_i\|}\cdot \sqrt{m_i}{\|\bm x_i\|}\right)^2\\
    &\leq& \sum_{i}  \left(\sqrt{m_i}\frac{{\bm x}_i\wedge {\bm x}_i}{ r_i}\right)^2 \sum_j \left(\sqrt{m_j}{r_j}\right)^2\\
\end{eqnarray}
We now use the fact that $\|\frac{{\bm r}_i\wedge {\bm v}_i}{r_i}\|^2\leq v_r^2$ and obtain
\begin{eqnarray}
    L^2\leq 2 E_t I
\end{eqnarray}
where $I=\sum_i r_i^2 m_i$ is the total moment of inertia.
Then, since $r_i^2<R^2$, we have $I\leq R^2 \sum_i m_i$.
Since $E=\sum_i (E_r^i+E_t^i)$ is the sum of the radial and tangential energy with respect to the radius of the particle $i$, we get $E_t\leq E$ and we obtain the final result that
\begin{eqnarray}
    L^2\leq 2 E_t I\leq 2 E (M R^2) \rightarrow \Xi=\frac{L^2}{2 M R^2 E}\leq 1
\end{eqnarray}
where $M=\sum_i m_i$ and $E=\sum_i \frac{{p}_i^2}{2m_i}$. This bound is saturated if all particles are at the maximum possible radius with kinetic energy along the radius, for which the moment of inertia is exactly $I=R^2 \sum_{i} m_i$ and $E^i=E_t^i$. We now discuss the equilibrium distribution using the MaxEnt principle.

\subsection{MaxEnt distribution}
First, let us mention that what we want to study is different from the physics of a gas in a rotating frame and that the statistical mechanics would be different. In the rotating frame, the Hamiltonian is replaced by 
    $H\rightarrow H-\vec \Omega \cdot \vec L$,
and thus the physical properties are different from the conservation of angular momentum \cite{LandauLif}. Although there are a lot of studies on rotating systems \cite{rot1,rot2}, we have not found in the literature an analysis of the implications of having a conserved angular momentum in the asymptotic distribution, and in particular the connection to the conserved order parameter $\Xi$ that we provide here. Although one possibility is to look at probability measures of the form $d\pi\propto e^{-\beta H} \delta(L_z-\sum_i {\bm p}\wedge {\bm x})$ \cite{dubrovskyi}, it is easier to look first at the MaxEnt principle to study the deviations from the Gibbs ensemble \cite{jaynes,jaynes2}, which is the simplest approach to understanding the issue in the presence of multiple conserved quantities \cite{Kinchin}.
 
The entropy of a classical statistical system in phase space is described, given a joint distribution $P(\bm x,\bm p)$, given by the maximization of
\begin{equation}
    S=-\int d^{N}{\bm x} d^{N}{\bm p}\ P(\bm x,\bm p)\log P(\bm x,\bm p)
\end{equation}
subject to the constraints
\begin{eqnarray}
    \int d^{N}{\bm x} d^{N}{\bm p}\ P(\bm x,\bm p)&=& 1\label{eq:probcons}\\
    \int d^{N}{\bm x} d^{N}{\bm p}\ P(\bm x,\bm p)E(\bm x,\bm p)&=& E_0\label{eq:encons}
\end{eqnarray}
In addition to the constraints above, however, in two dimensions we need to add the constraint
\begin{eqnarray}
        \int d^{N}{\bm x} d^{N}{\bm p}\ P(\bm x,\bm p)L_z(\bm x,\bm p)&=& L_0\label{eq:momcons}
\end{eqnarray}
for the angular momentum. We introduce the Lagrange multipliers $\beta_P,\beta_T$ and $\beta_L$ for the probability conservation, energy conservation, and momentum conservation respectively, it is not hard to see that one obtains
\begin{eqnarray}
    P(\bm x,\bm p)    &=&\prod_{i=1}^N \frac{1}{\mathcal N_1}e^{-\frac{\beta_T {\bm p_i}^2}{2 m}-\beta_L ({\bm p}_i\wedge {\bm x}_i)_z }
\end{eqnarray}
which is written in terms of factorized one-body terms.  The Lagrange multiplier, unlike the $\beta_T=\frac{1}{\kappa_B T}$ which is a unit of inverse energy, is in inverse units of angular momentum. The first comment we wish to make is that the distribution is not time-reversal invariant, as $\mathcal T(L_z)=-L_z$, where $\mathcal T(p)=-p$, $\mathcal T(x)=x$ is the time reversal operation. This implies that when angular momentum is conserved, the equilibrium distributions are different if time flows forward or backward. This is an important departure from Gibbs distributions. The derivations of the formulae shown below are provided in App. \ref{app:maxentprinc}.

\textbf{Energy distribution in space}. Focusing on the one body distribution:
\begin{eqnarray}
    P_1({\bm p}, {\bm x})&&=\frac{1}{\mathcal N_1}e^{-\beta_T \frac{p_x^2+p_y^2}{2m}-\beta_L(p_y x-p_x y)}\nonumber \\
    &=&\frac{1}{\mathcal N_1} e^{-\beta_T \frac{p_x^2}{2m}+\beta_L p_x y}e^{-\beta_T \frac{p_y^2}{2m}-\beta_L p_y x}
\end{eqnarray}
from which it is easy to see that the distribution is factorized in $p_x$ and $p_y$.  We have in particular that, if we condition the one-body distribution on the position, using 
\begin{eqnarray}
    P_1({\bm p}| {\bm x})=\frac{P_1({\bm p}, {\bm x})}{\int d^2p P_1({\bm p}, {\bm x})},
\end{eqnarray}
we have
\begin{eqnarray}
   \frac{m \left(\frac{\beta_L^2}{2} y^2 m+\beta_T \right)}{\beta_T ^2}= \langle p_x^2\rangle({\bm x}) \neq \langle p_y^2\rangle({\bm x})=\frac{m \left(\frac{\beta_L^2}{2} x^2 m+\beta_T \right)}{\beta_T ^2}.\nonumber 
\end{eqnarray}
The equality is restored in the limit $\beta_L\rightarrow 0$.
It follows that the energy density is concentrated at the boundary, e.g.
\begin{eqnarray}
    \langle E\rangle({\bm x})=\langle \frac{p_x^2+p_y^2}{2m} \rangle({\bm x})=\frac{\frac{\beta_L^2}{2} (x^2+y^2) m+2\beta_T}{2\beta_T^2}.
\end{eqnarray}
This is one of the interesting features of having angular momentum in the system, and we will see in numerical simulations that energy is localized.

\textbf{Condensation phenomenon}. From now on, we switch to adimensional momentum and position coordinates. 
Specifically, the adimensional 1-body distribution we consider is equivalent to the following definition in terms of the adimensional scalars  $ p\in[0,\infty)$ and $ r\in[0,1]$ as
\begin{eqnarray}
     P_1(p,r)&=&\Large \langle \frac{1}{N}\sum_{i=1}^N\delta(\frac{\sqrt{\beta_T}
     }{\sqrt{m}} p-\|\bm p_i\|)\delta(R r-\|\bm x_i\|)\Large \rangle_{P(\bm p,\bm x)}.\nonumber
\end{eqnarray}

As we show in App. \ref{app:maxentprinc}, the one body distributions in momentum and position, with respect to the dimensionless 
\begin{eqnarray}
    P_1(p,r)&=&\frac{1}{\mathcal N}\ p\ r\ e^{-\frac{p^2}{2} }I_0(\kappa p r),\ \ \ \ \mathcal N=\frac{1}{\pi}\frac{e^{\gamma}-1}{\gamma}\nonumber \\
    P_1^\gamma(r)&=&\int_0^\infty dp P_1(p,r)=\frac{2 \gamma  r e^{\gamma  r^2}}{e^{\gamma }-1}\label{eq:p1gammar}\\
    P_1^\gamma(p)&=&\int_0^\infty dr P_1(p,r)= \frac{\gamma  e^{-\frac{p^2}{2}} p \, _0\tilde{F}_1\left(2;\frac{p^2 \gamma }{2}\right)}{e^{\gamma }-1}\nonumber \\
    \kappa &=&\frac{\beta_L R \sqrt{m}}{\sqrt{\beta_T}}\label{eq:p1gammap}
\end{eqnarray}
where $I_0(x)$ is the modified Bessel function of the first kind for $n=0$, $F_1(2;x)$ is  the regularized confluent hypergeometric function, and  where we defined
\begin{equation}
    \gamma=\frac{\kappa^2}{2}=\frac{1}{2}\Big(\frac{\beta_L R \sqrt{m}}{\sqrt{\beta_T}}\Big)^2.
\end{equation}
As we show in App. \ref{sec:chiparams}, $\gamma$ is a continuous and monotonic function of $\Xi$. In particular,  $\lim_{\Xi\rightarrow 0} \gamma(\Xi)=0$ and $\lim_{\Xi\rightarrow 1} \gamma(\Xi)=\infty$. This shows why the order parameter $\Xi$ characterizes the behavior of the system depending on whether the angular momentum is conserved or not, as shown in Fig. \ref{fig:summary}.

\begin{figure}
    \includegraphics[width=\linewidth]{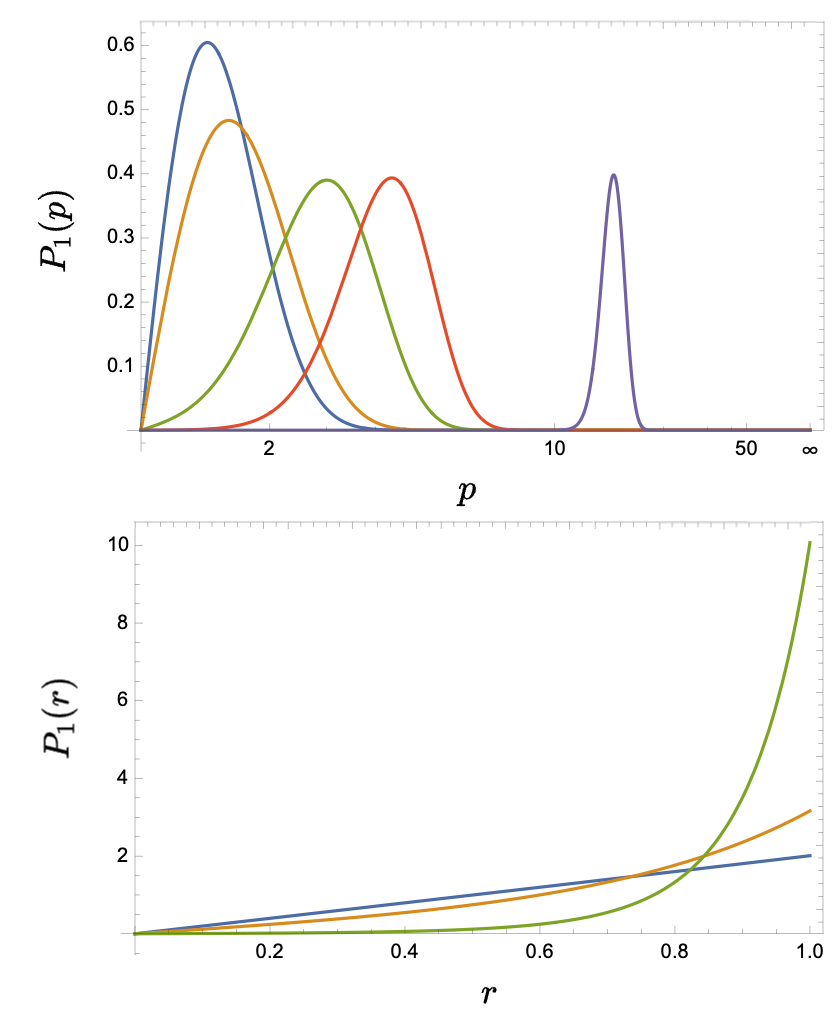}\\
    \caption{Distributions $P_1^\gamma(p)$ (top) and $P_1^\gamma(r)$ (bottom) as a function of $\gamma=0,1,10,100,1000$ and $\gamma=0,1,10,100$ respectively. We see from the 1-body distributions above that for $\gamma\gg 1$ we have a condensation phenomenon around the boundary.}
    \label{fig:distsmp}
\end{figure}

The behavior of these two distributions as a function of $\gamma$
can be seen in Fig. \ref{fig:distsmp}.
For the joint one-body distribution
in particular, we have
\begin{eqnarray}
    \langle  E \rangle&=&\langle \frac{u^2}{2}\rangle=\left(\frac{1}{e^{\gamma }-1}+1\right) \gamma\\
\langle  L \rangle&=&\langle p r \sin \alpha\rangle=-\frac{\sqrt{2} \left(e^{\gamma } (\gamma -1)+1\right)}{\left(e^{\gamma }-1\right) \sqrt{\gamma }}
\end{eqnarray}
Expressions for the limits $\gamma\rightarrow 0$ and $\gamma\rightarrow \infty$ can be evaluated analytically. 
In particular, for $\Xi=1$, we have that the radial distribution becomes a Dirac $\delta$-function centered on $r=1$, e.g.
\begin{eqnarray}
    \lim_{\gamma \rightarrow \infty} P_1^\gamma(r)=2\delta(1-r)\label{eq:condensation}
\end{eqnarray}
consistent with numerical simulations. We interpreted this behavior as a condensation phenomenon near the boundary.  In the same limit,  the momentum distribution peaks at larger and larger momenta, also consistent with numerical observations. A way to interpret this condensation is in terms of a single and non-interacting particle, with an initial position near the boundary and with speed perpendicular to the radius, e.g. with $\Xi\approx 1$. Since the collision with the boundary preserves the angle of incidence \cite{circ1}, a particle starting in this condition will remain near the boundary since the incidence angle is approximately zero. This condensation phenomenon is reminiscent of this single-particle behavior, with the caveat that collisions are an important factor that is included in our distribution.


\textbf{Canonical partition function and pressure}. To conclude this section, let us discuss the pressure of such a gas as compared to an ideal gas.
To calculate the pressure, we use the formula $P=-\partial_V F$, with $F=-\frac{1}{\beta_T} \log Z$. The partition function is, up to constants, given by (see App. \ref{app:maxentprinc}):
\begin{eqnarray}
    Z_N\propto \left(\frac{ e^{\frac{m R^2 \beta _L^2}{2 \beta _T}}-1}{\beta _L^2}\right)^N
\end{eqnarray}
from which, using the fact that $V=\pi R^2$, we obtain
\begin{eqnarray}
    P= \frac{m N \beta _L^2}{2 \pi  \beta _T^2 \left(1-e^{-\frac{m V \beta _L^2}{2 \pi  \beta _T}}\right)}
\end{eqnarray}

In the limit $\beta_L\rightarrow 0$, we recover $PV=N k_B T$. At the first non-zero order correction, we have instead
\begin{eqnarray}
   PV=\frac{N}{\beta_T}\Big(1+V\frac{m  \beta _L^2}{4 \pi  \beta _T}\Big)+O(\beta_L^4)
\end{eqnarray}
which shows that the ideal gas law is modified by the conservation of angular momentum. While at low temperatures this correction is less important, it becomes more important at larger temperatures. Also, we note that the correction is always positive. 
For the pressure, we have the correction from the gas
\begin{eqnarray}
    \frac{P-P_{ideal}}{N}=\frac{m \beta_L^2}{4 \pi \beta_T^2} +O(\beta_L^4)
\end{eqnarray}
which is \textit{independent from the volume}.

This implies that in the presence of angular momentum and at equal volume and temperature, the pressure of a fluid with more angular momentum is larger than one at lower angular momentum. It is also important to stress that this contribution is not due to the virial expansion, as it does not incorporate the finite size of the particles. It is a genuine contribution due to the fact that our measure is not given by a Gibbs distribution. A physical intuition of this fact can be obtained by the observation that particles spend more time near the boundary than in the case of an ideal gas. This is a peculiar type of condensation phenomenon, and it depends on the value of the order parameter $\Xi$ that the gas attains initially.

\begin{figure*}
\includegraphics[width=0.75\linewidth]{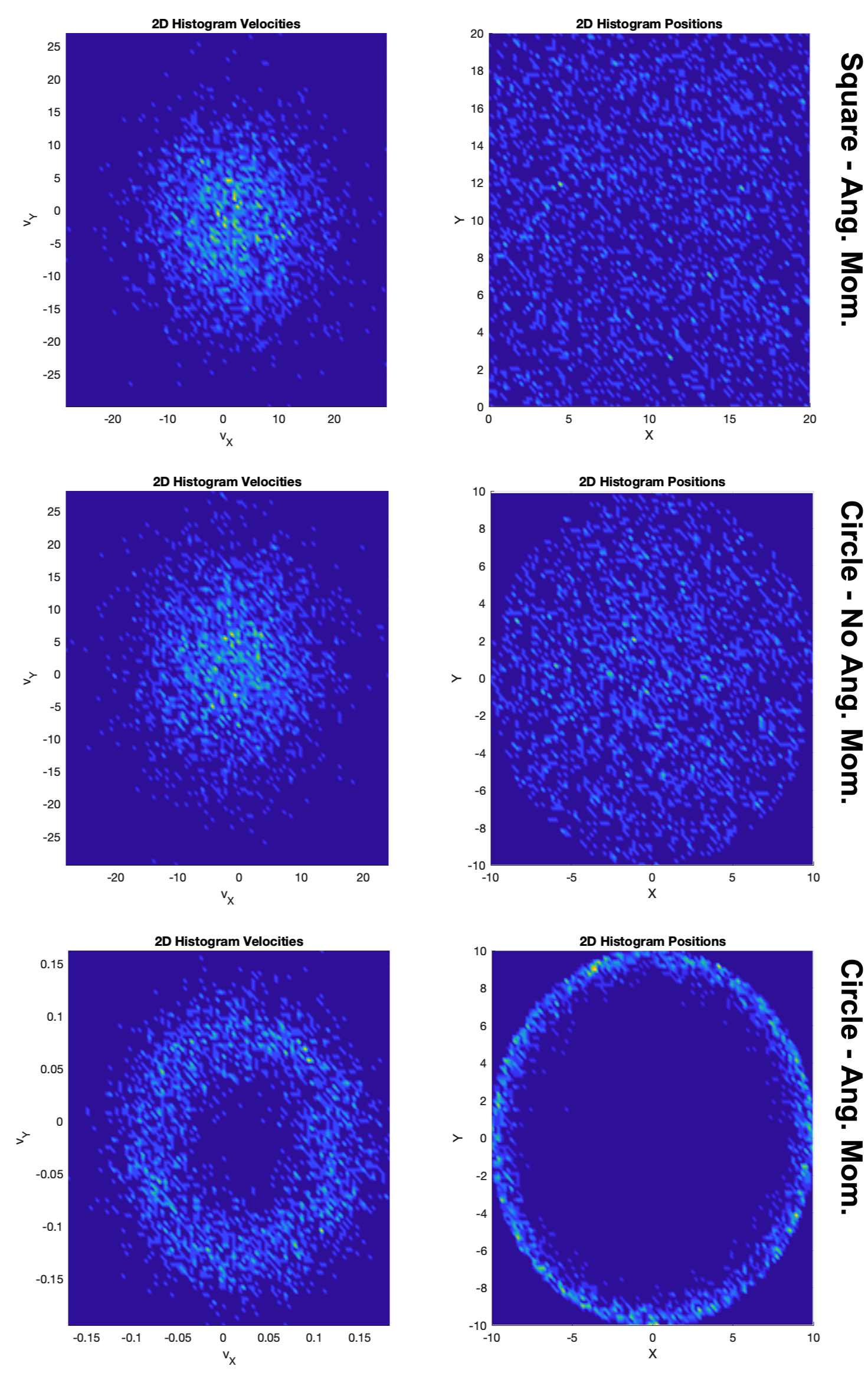}
\caption{EDMD simulations of particle collisions for the square, and circle, with 2500 particles and a maximum of 20000 collisions, with packing fraction $\eta=0.1$. With initial angular momentum, $\Xi\approx 0.6$ (top), the square has asymptotic 1-body distributions consistent with Gibbs, e.g. uniform spatially and Gibbs distributed in the velocity. We see the same results for the circle, with no angular momentum in the initial state, $\Xi\approx 0$ (center plots). Instead, if the initial state has angular momentum, $\Xi\approx 0.6$ and is identical to the top initial state, the system's asymptotic distribution exhibits a condensation phenomenon around the boundary, predicted by the MaxEnt distribution. Similar results, with $6000$ particles and higher initial speeds, were obtained with TDMD simulations (see App. \ref{ref:clust}).}
\label{fig:eventdriven}
\end{figure*}


\section{Simulations}
The theoretical model we have introduced in the previous section predicts that the condensation phenomenon, as a result of having initial states with high angular momentum, should be predicted by the initial value attained by the order parameter $\Xi$. It is worth mentioning that this type of angular momentum is not induced by the chirality of particles for a granular gas with dissipative forces, as for instance observed in \cite{gollub}. It is a property of the initial state, and because of ergodicity breaking, the system will be trapped in it if angular momentum is conserved alongside energy. The deviation from the Gibbs ensemble should then be observable in simple models, as long as angular momentum and energy are both conserved. 

One of the simplest models to study thermalization, ever since Boltzmann's introduction of the H-theorem and the BGGKY hierarchy \cite{huang}, is the hard disks model. The model is described in simple terms: particles, which are disks of finite radius, interact through a purely repulsive potential infinitely short-ranged, resulting in an impenetrable core. In the absence of long-range interactions between the particles, this model can be simulated using an Event Driven Molecular Dynamics simulation, introduced in \cite{lanl3}. The model considers rigid objects of finite area, essentially two-dimensional billiard balls. These are particles that, in principle, because of the hardcore potential could not be simulated using time-driven simulations because of the infinite forces. However, as we show in the App. \ref{ref:clust}, a similar phenomenology as the one we show here is also observed in a naive TDMD. One of the advantages (but at the same time a disadvantage for more advanced applications) of EDMD is that no forces need to be calculated, and one introduces no arbitrary, unphysical, time steps as compared to TDMD.   This implies that the particles instantaneously exchange their momenta perpendicular to the line joining their centers ($\mathbf{r}_{ij}$) while conserving their total momentum. 

\subsection{EDMD setup}
 As is typically the case in naive EDMD like the one we consider, one needs to calculate the time $t_c$ between collisions and create a lookup table.  Collisions are only of two types: particle-particle and particle-boundary.  In our EDMD, particles follow straight lines in between collisions, which are assumed to be instantaneous events (See for instance \cite{Haile}, Sec. 3.2.1).   

\textbf{Particle collisions.} There are no interactions between particles in between collisions, e.g. there are no long-range interactions. Each particle is described by its position and velocity (or momentum for fixed mass) vectors $(\mathbf{r}_i, \mathbf{v}_i)$. Collisions are instantaneous and induce no deformations on the disks, and both momentum and energy are conserved. The interparticle potential as a function of the distance between the particles, $r_{ij} = |\mathbf{r}_i - \mathbf{r}_j|$, can be assumed to be infinitely short-ranged:
$$
U(r_{ij}) = \begin{cases}
\infty & \text{if } r_{ij} < d \\
0 & \text{if } r_{ij} \geq d
\end{cases}
$$
where $r_{ij}$ is the distance between the particle centers and $d$ is the diameter of the hard disks. This potential ensures that particles cannot approach each other closer than $d$ units. Thus, time is evolved until the moment particles touch. At that moment, the velocity between the particles is updated.
In particular, collisions between two hard disks $i$ and $j$ conserve momentum:
$$
m_i \mathbf{v}_i' + m_j \mathbf{v}_j' = m_i \mathbf{v}_i + m_j \mathbf{v}_j
$$
where $m_i$ and $m_j$ are the masses, and $\mathbf{v}_i$, $\mathbf{v}_j$ are velocities before the collision. The relative velocity between the particles along the line of centers ($\mathbf{v}_{ij}$) is important:
$$
\mathbf{v}_{ij} = \mathbf{v}_i - \mathbf{v}_j
$$
The post-collision velocities can be found using the formulas for elastic collisions:
\begin{eqnarray}
\mathbf{v}_i' &=& \mathbf{v}_i - \frac{2 m_j}{m_i + m_j} \left(\mathbf{v}_{ij} \cdot \hat{\mathbf{r}}_{ij}\right) \hat{\mathbf{r}}_{ij} \\
\mathbf{v}_j' &=& \mathbf{v}_j + \frac{2 m_i}{m_i + m_j} \left(\mathbf{v}_{ij} \cdot \hat{\mathbf{r}}_{ij}\right) \hat{\mathbf{r}}_{ij}
\end{eqnarray}
where $\hat{\mathbf{r}}_{ij}$ is the unit vector along the line of centers. In our simulations, $m_i=1$ and thus the equations above are further simplified.  As we can see, since the conservation of angular momentum that we study is a property of the collision between spherical/circular disks and the circular boundary, we do not consider any complicated interactions between the disks \cite{shafer} such as friction or intrinsic particles angular momentum changes, e.g. particles are not rotating around their center of mass.
Time is evolved until particles touch, and the momentum is updated at that event. We note that in collisions between hard disks, the total angular momentum
$$
\vec{L} = \sum_i \vec{l}_i = \sum_i m_i \mathbf{v}_i \wedge \mathbf{r}_{\text{ref}}
$$
where $\mathbf{r}_{\text{ref}}$ is any fixed vector is also conserved. In two dimensions, this is a scalar corresponding to the $z$-component of the total vector, $L_z = \sum_i m_i (\mathbf{p}_i \wedge \mathbf{r}_{\text{ref}})_z$.

\textbf{Boundary collisions}. Let us now discuss the collision with the boundary. When a particle reaches a  boundary, its velocity vector $\mathbf{v}$ undergoes reflection relative to the local normal vector $\hat{\mathbf{n}}$ at the point of impact:
$$
\mathbf{v}' = \mathbf{v} - 2(\mathbf{v} \cdot \hat{\mathbf{n}})\hat{\mathbf{n}}
$$
where $\hat{\mathbf{n}}$ at the point of impact $(x, y)$ can be computed as:
$$
\hat{\mathbf{n}} = \frac{(x, y)}{\sqrt{x^2 + y^2}}
$$
Since we are also working with circular boundaries, the only difference with standard simulations is the calculation of the time of collision, which we report for completeness in the App. \ref{sec:edmd}.

 For a circular boundary, the radius always aligns with the tangent surface, preserving angular momentum. Alternatively, $\hat{\mathbf{n}}=\hat{\mathbf{r}}$. Thus, choosing $\mathbf{r}_{\text{ref}}=\hat{\mathbf{r}}$, implies that, for a single particle $i$ at collision with the boundary, we have  
 \begin{eqnarray}
     L_i^\prime \hat{\mathbf{z}}= m \mathbf{v}'_i \wedge \hat{\mathbf{r}}= m \mathbf{v}_i \wedge \hat{\mathbf{r}}= L_i \hat{\mathbf{z}}
 \end{eqnarray}
where we used $\hat{\mathbf{r}}\wedge \hat{\mathbf{r}}=0$, and thus each particle will conserve its own angular momentum with respect to the center of the circular boundary.
 
 This is unique to circular boundaries, where the initial angular momentum:
$$
L_0 \hat{\mathbf{z}} = \sum_{i} (p_y^i x_i - p_x^i y_i)\hat{\mathbf{z}} = \sum_i \mathbf{p}_i \wedge \mathbf{x}_i
$$
is conserved. 
This paper investigates this case, highlighting strong ergodicity breaking in a microcanonical configuration depending on initial conditions based on boundary effects only. We focus on the $\hat{z}$ component of angular momentum in two dimensions.

\subsection{Numerical results}

To begin with, let us explain how the initial conditions were chosen. The initial conditions were chosen at random, but consistent with the constraint that they have specific angular momentum and a particular value of $\Xi$.  In the limit in which particles are diluted, $\eta=0.1$, we consider an initial uniform spatial distribution confined to an annulus centered around $\tilde r=g R$, with $0\leq g\leq 1$ and of thickness $\delta$. Each particle's momentum is oriented according to an angle $\theta$, and with well-specified momentum $m v_0$. If the angle $\theta$ is distributed uniformly in $[0,2\pi]$, it is easy to see that the total angular momentum is on average zero in this configuration. If instead $\theta$ is chosen to be oriented parallel to the tangent of the annulus, then $|L_z|\geq 0$. In particular, one can see that $\Xi \approx g^2$, and thus we can control $\Xi$ by choosing the initial condition's initial radial distribution to be thin and centered at a value close to the boundary.

In the case of the event-driven simulations, for the case of the square boundary, we have used the same methodology as in Haile (see Sec. 3.2.1) \cite{Haile}. In event-driven simulation, particles move in straight-line trajectories at constant speeds between collisions. The simulation focuses on predicting and processing collision events.  As common, we maintain a priority queue of future collision events ordered by time and advance the system to the time of the next event.  The only difference with EDMD simulations is the calculation of the collision time with the circular boundary, reported in the App. \ref{sec:edmd}. We then update particle velocities according to collision rules.   This method is efficient as it processes only significant events, and it simulates real hard disks at low packing fractions, reducing unnecessary calculations between collisions.  For a number of $2500$ disks and after 20000 collisions, the results are shown in Fig. \ref{fig:eventdriven}.  As we can see, in the case of square boundaries, as expected, even in the case with $\Xi\approx 0.6$, the momentum distribution reduces to a Gibbs distribution centered in zero, while the position distribution is uniform over the area. The same occurs for the circular boundary in the absence of initial angular momentum. However, when $\Xi\approx 0.6$, the asymptotic distribution shows deviations from the Gibbs distribution consistent with the analytical results from the previous section. In this case, the particles exhibit the condensation phenomenon near the boundary.

We obtained analogous results in the case of TDMD simulations, where however it could be argued that these simulations are not hard disks with infinite potential as there is a small overlap of order $dt$. These results are shown in the App. \ref{ref:clust}, with 6000 disks, with simulations ran on a cluster. We obtain an analogous condensation phenomenon and dependence on the initial configuration's order parameter $\Xi$. We simply interpret these results as the robustness of the deviations from the Gibbs asymptotic distributions to details of the interactions between the particles. One of the advantages of using this method is that we can see the time evolution of the order parameter $\Xi(t)$ as a function of time, due to the collisions with a boundary not perfectly circular, and in particular as a function of the eccentricity for elliptical boundaries. As expected, the parameter $\Xi$ decays significantly over time for larger eccentricity.

\section{Applications}
\subsection{Modified Metropolis-Hastings algorithm}

\begin{figure*}
\includegraphics[width=0.99\textwidth]{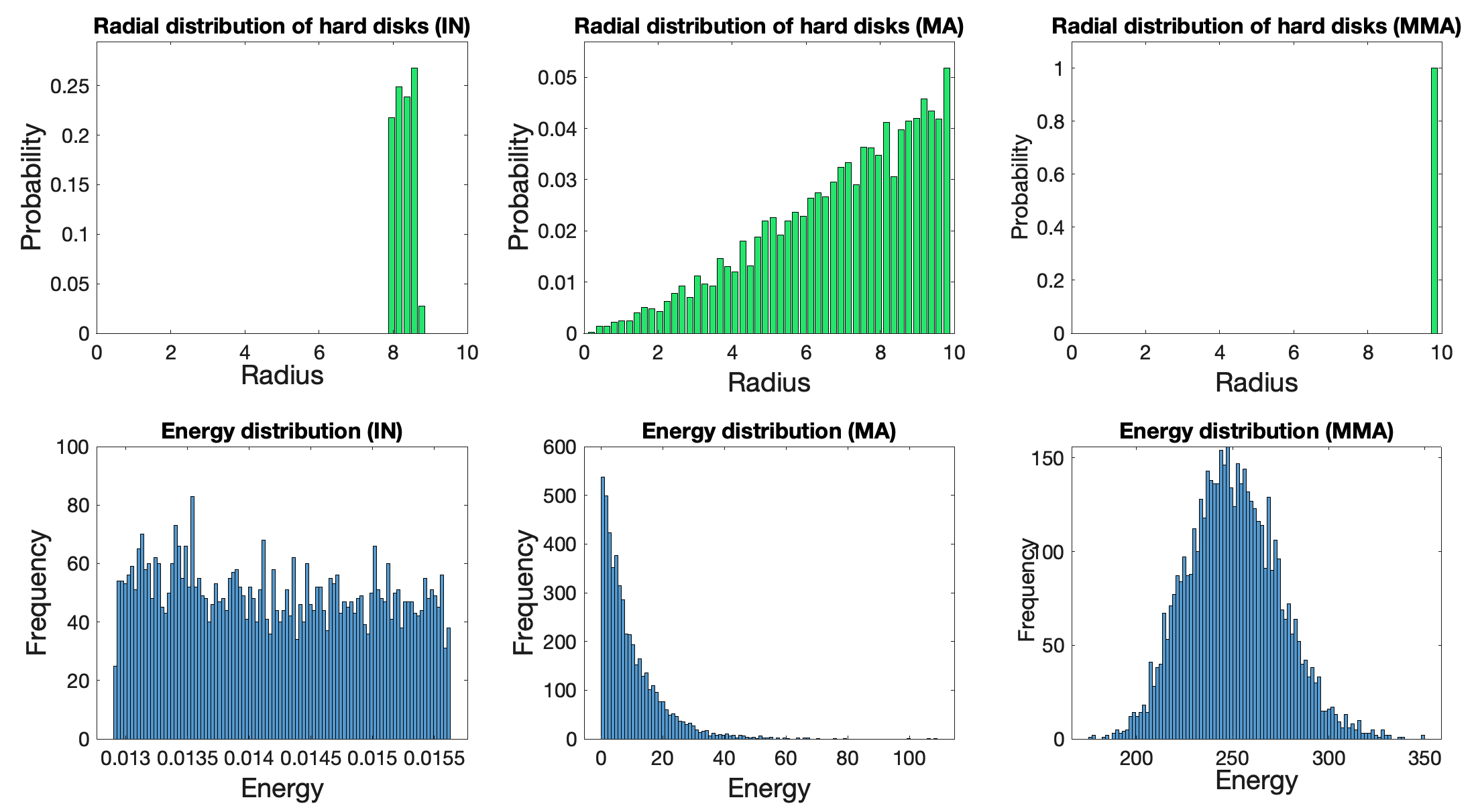}
\caption{Above, we present graphs of the probability of particles being at a certain radius $r$, in a circular boundary with radius $R=10$. In the simulations above, we have a relatively small number of particles, $N_p=5000$, and the simulation is run for $N_s=10^7$ steps. The parameters of the MCMC are the mean initial position of the graph $r_{in}=0.85$, and the frequency of particles at a certain energy, on the two graphs on the left (IN, top and bottom respectively). We consider two MCMCs. The first is the standard MCMC algorithm, which we refer to as MA following eqn. (\ref{eq:ma}), with the distribution of the particles' initial condition being those on the left, based only on energy and allowed positions, with $\beta_T^{-1}=100$, $m=1$. 
The results after $N_s$ steps are shown in the center panels for the radial probability (MA, center, top) and the frequency of energy (MA, center, bottom). The MMA algorithm uses the same parameters and initial conditions, with the addition of $\beta_L^{-1}=0.1$, and follows the acceptance probability of eqn. (\ref{eq:mma}). The results after $N_s$ steps are shown on the right panels for the radial distribution (top) and energy (bottom). We can see the condensation phenomenon near the boundary on the radial distribution, and the non-centered in zero distribution of energy showing a deviation from Gibbs.}
\label{fig:mc}
\end{figure*}

Let us briefly discuss why these distributions cannot be obtained via the standard Metropolis-Hasting algorithm.
Originating from the seminal work of Metropolis et al. and generalized by Hastings  \cite{metropolis1953equation,hastings1970monte}, this algorithm is used to understand high-dimensional configuration spaces, making it indispensable for studying systems like hard disks. In the context of hard disk systems, the Metropolis-Hastings algorithm enables the generation of configurations representative of the system's equilibrium state. This process involves iteratively proposing and accepting or rejecting moves that modify the positions of the disks while preserving key physical constraints, such as excluded volume interactions and boundary conditions. At each iteration of the algorithm, a new configuration is proposed by randomly selecting a disk and attempting to move it within the system. The acceptance or rejection of this move is determined by evaluating an acceptance probability, and the formula used for the accepting/rejection is  \cite{landaubinder}
\begin{equation}
    P_{acc}=\textit{min}(e^{-\beta (E_{new}-E_{old})},1)\label{eq:ma}.
\end{equation}
We refer to this algorithm as MA (Metropolis Algorithm).
The issue with this formula is that it assumes that the equilibrium distribution is the Gibbs ensemble $P_{eq}\propto e^{-\beta E}$. Thus, using this type of Monte Carlo scheme, as we have assessed so far, would provide the wrong distribution in the case of the circle for the hard disks. Unsurprisingly, it is possible to modify the acceptance probability to account for the conservation of angular momentum $L_z$:
\begin{equation}
    P_{acc}=\textit{min}(e^{-\beta_T (E_{new}-E_{old})-\beta_L (L_{new}-L_{old})},1).\label{eq:mma}
\end{equation}
We will refer to this algorithm as Modified Metropolis Algorithm, MMA. 
We compare these two algorithms for the circular boundaries in Fig. \ref{fig:mc}. Given the dynamical evolution of the particles, we see the use of the acceptance probability in eqn. (\ref{eq:ma}) is not correct in this case, and instead the acceptance probability of eqn. (\ref{eq:mma}) is the right choice, providing results that are consistent with eqn. (\ref{eq:p1gammar}) and eqn. (\ref{eq:p1gammap}) for $\gamma=\frac{1}{2}(\frac{10\cdot 10}{\sqrt{100}})^2\approx 10^2$, thus in the regime in which we expect strong deviations from the Gibbs distributions. We see instead that the MA algorithm converges to the Gibbs distribution (as designed). This is, in one way, the drawback of choosing from the beginning the distribution to which the system should converge, and which however has to be tied to the dynamics of the system. This is one example where this choice is important. Although the relative number of particles is small compared to the state of the art, our simulations show sufficiently different results that these details should not be important. It is important to stress, however, that we are far away from $\eta_c\approx0.7$, with the packing $\eta\approx 0.1$, and thus using the right distribution might be important in this case, and modify dramatically the nature of the transition. 
\subsection{Deviations from the Bohr-van Leeuwen theorem}
As curious application of the results of this manuscript, let us consider a topic of historical importance, such as the Bohr–van Leeuwen theorem. Bohr discovered this theorem in his doctoral thesis and was later independently identified by van Leeuwen in her 1919 doctoral thesis, addressing the classicality of electric and magnetic susceptibilities \cite{bohr,vl,vv}. The theorem's historical significance lies in its challenging classical physics, by revealing its inability to account for phenomena like paramagnetism, diamagnetism, and ferromagnetism. This inadequacy necessitated the introduction of quantum physics to explain magnetic occurrences. This theorem might have influenced Bohr's development of a quasi-classical theory of the hydrogen atom in 1913 and its importance must not be undermined. However, we stress that this theorem is based on the assumption of an asymptotic Gibbs distribution. 

The proof typically goes as follows. Consider a classical system of charged particles in thermal equilibrium subjected to conservative forces.
The magnetic moment ${\bm m}_i$ for a charged particle with charge $e$, velocity $\mathbf{v}_i$, and mass $m_i$ is given by:
\begin{equation}
{\bm m}_i = \frac{e}{2m_i} {\bm v}_i \wedge {\bm r}_i
\end{equation}
where $\mathbf{r}_i$ is the position vector of the particle. The total magnetization ${\bm m}$ of the system is the sum of these magnetic moments for each particle ${\bm m}_i$:
\begin{equation}
{\bm m} = \sum_i {\bm m}_i
\end{equation}

In thermal equilibrium, the Maxwell-Boltzmann distribution describes the velocity distribution of particles:
\begin{equation}
f({\bm v}) \propto e^{- \frac{\beta_T}{2} m v^2}.
\end{equation}

The average magnetic moment for a single particle is:
\begin{equation}
\langle {\bm m}_i \rangle = \int {\bm m}_i f(\mathbf{v}_i) d\mathbf{v}_i = \frac{e}{2m} \langle {\bm v}_i \wedge {\bm r}_i \rangle = 0
\end{equation}
since the average of ${\bm v}_i$ and $\vec {r}_i$ are zero due to isotropy.

Thus, the total magnetization $\mathbf{M}$ is the sum of zero-mean random variables, resulting in:
\begin{equation}
\langle {\bm m} \rangle = \sum_i \langle{\bm m}  _i \rangle = 0
\end{equation}

Therefore, in thermal equilibrium, the average magnetization of the system is zero, confirming the absence of spontaneous magnetization in classical systems under conservative forces. This is the essential statement of the Bohr-van Leeuwen theorem.

However, when angular momentum is conserved, we find that such quantity is no longer zero. This is connected to the fact that our distribution is not invariant under time reversal. This was also mentioned in~\cite{drubovskyi2}. In particular, we find that for our two-dimensional distributions for which ${\bm m}$ is a scalar, we have
\begin{eqnarray}
\frac{\langle  M \rangle}{N} &=&
\frac{2}{\beta _L}-\frac{m R^2 \beta _L \left(\coth \left(\frac{m R^2 \beta _L^2}{4 \beta _T}\right)+1\right)}{2 \beta _T}
\end{eqnarray}
For small values of $\beta_L$, we get 
\begin{eqnarray}
    \frac{\langle  M \rangle}{N} &\approx & -\frac{e R^2 \beta _L}{4 \beta _T}+O(\beta_L^3)
\end{eqnarray}
where the assumption is that $\beta_L^2 << \frac{\beta_T}{m R^2}$. 
Thus, we recover the Bohr-van Leeuwen theorem in the limit in which the conservation of angular momentum is less important. From a physical perspective, a perfect boundary is highly unlikely to happen, and we expect the persistence of magnetic fields in certain metallic materials with a high degree of symmetry. Paradoxically, the violation of time reversal symmetry is exactly what led to the discovery of the electrons, as correctly pointed out in \cite{drubovskyi2}. We refer to the Tolman-Stewart experiment, which provided the first evidence for the existence of electrons in a conductor \cite{tolman}. In this experiment, a conductor coil is set up with a certain angular momentum due to its rotation. When the rotation is suddenly halted, an electric field is induced in the coil. This induced electric field is a consequence of Faraday's law of electromagnetic induction, which states that a changing magnetic field induces an electric field. As a result, the angular momentum of the coil (and its electron gas) induced a magnetic field.

\section{Conclusion}

The present manuscript investigated the relaxation to an equilibrium distribution of a system of hard disks, revealing a notable deviation from the expected Gibbs distribution in the case of the circle (and near-circle) boundaries. 
This difference from the anticipated statistical behavior is a crucial departure from classical predictions in statistical mechanics. In particular, we have shown that in this case another physical quantity is conserved, in particular the total angular momentum.
By examining the dynamics and interactions among hard disks, the study reveals that the initial conditions in reality determine the asymptotic distribution. This is a deviation from the assumption that the initial conditions are not important in the thermalization process. 

The exploration into the deviation from Gibbs distribution in the hard disk system was approached using  Jayne's maximum entropy principle. This principle, led to the formulation of the Generalized Gibbs ensemble as the statistical framework describing the observed distribution of hard disks and the limits, in terms of the order parameter we introduced in this manuscript, of low and high initial angular momenta.

The emergence of the Generalized Gibbs ensemble underscores its significance in capturing the intricate dynamics and interactions within the hard disk system that lead to the observed deviations. 
Moreover, the utilization of the maximum entropy principle to derive the Generalized Gibbs ensemble not only provides a statistical foundation for understanding the observed deviations but also sheds light on the applicability and limitations of traditional statistical mechanics in describing complex classical systems, even if as simple as hard disks. This begs the question of why this has not been observed earlier. One possibility is how initial conditions are provided. Essentially, we show that there is a hard ergodicity breaking. If the particles are initiated at random, angular momentum will likely be zero on average. The system will remain in this zero angular momentum state, remembering its initial condition. This state is determined by the order parameter $\Xi$, as we have shown.

Also, we have considered one possible perturbation of the circular boundary geometry, namely the ellipse. Different ways of perturbing the circular boundary might lead to peculiar phenomena of weak ergodicity breaking. Given the widespread applicability of the hard sphere model, we think that this point is not stressed enough in the literature. One of the criticisms and limitations of this paper is fine-tuned to a specific boundary, which as we have shown in Fig. \ref{fig:eccchit} limits the applicability of our approach. However, it is easy to see that this result is also valid in the absence of boundaries, when the system is confined by a central potential. This could be, for instance, a large mass,  at the center of the circular boundary, interacting with all the particles via pairwise potentials. This will be analyzed in detail in future studies. The second limitation of our paper is that it is focused only on the liquid phase of the hard disk model. This phase requires more sophisticated event-driven MC methods, and we think it could be an area of interest for experts in the field, as we think that the MMA will be an important factor to consider for circular boundaries and that it could modify the nature of the transition and the critical values $\eta_c$.

The findings not only challenge the conventional assumptions of the Gibbs distribution but also establish a direct link to the violation of the Bohr-van Leeuwen theorem. This classical theorem postulates that in a purely classical system, thermal motion alone cannot generate magnetism. However, the observed deviation in the hard disk system suggests that certain classical charged systems with a high degree of symmetry, particularly those with highly constrained or intricate interactions, can indeed exhibit magnetic properties due to deviations from expected statistical distributions. 

The breakdown implies that certain classical systems, contrary to previous understanding, might exhibit intrinsic magnetic properties such as permanent magnets. This challenges the traditional viewpoint that thermal motion alone cannot generate magnetism in classical systems, indicating the potential for classical systems with specific characteristics to display magnetic behavior. In particular, this hints towards the persistence of induced magnetization in finely designed certain paramagnets.

Of course, there are various possible future directions. The first obvious generalization is to extend this result to higher dimensions. In the case of the sphere boundary with hard spheres, this result generalizes with quantity conserved being the total angular momentum $\vec L=\sum_i {\bm p}_i \wedge {\bm x}_i$. In higher dimensions, there are also some situations in which some components of the angular momentum are conserved. This is for instance the case for the cylinder in which only the component of the angular momentum along the long axis is conserved. In addition, in future studies, we will perform the calculation of the thermodynamic quantities such as pressure in the microcanonical ensemble.

\begin{acknowledgments}
Our work was conducted under the auspices of the National Nuclear Security Administration of the United States Department of Energy at Los Alamos National Laboratory (LANL) under Contract No. DE-AC52-06NA25396.  This research used resources provided by the Darwin testbed at Los Alamos National Laboratory (LANL) which is funded by the Computational Systems and Software Environments subprogram of LANL's Advanced Simulation and Computing program (NNSA/DOE). We are in particular indebted to  C. Nisoli, C. Schimming for their comments on the manuscripts and suggestions, Ł. Cincio for help in setting up the cluster code,   C. J. O. Reichhardt and C. Reichhardt for pointing out some drawbacks in our numerical methods of an earlier draft, and D. P. Landau and E. Ben Naim for various comments on the relevance of boundaries in numerical methods.
\end{acknowledgments}

\clearpage

\textbf{\Large APPENDIX}
\appendix
\section{The MaxEnt principle and canonical partition function} \label{app:maxentprinc}
In this Section, we derive the correction to the Maxwell-Boltzmann distribution from the MaxEnt principle. We consider the one-particle distribution obtained from 
\begin{eqnarray}
    P_1(p,r)&=&\Large \langle \frac{1}{N}\sum_{i=1}^N\delta(\sqrt{\frac{\beta_T}{m}}p-\|\bm p_i\|)\delta( R r-\|\bm x_i\|)\Large \rangle_{P(\bm p,\bm x)}\nonumber \\
    &=&\Large \langle \delta(\sqrt{\frac{\beta_T}{m}}p-\|\bm p_i\|)\delta( R r-\|\bm x_i\|)\Large \rangle_{p(\bm p_i,\bm x_i)}
\end{eqnarray}
where in the second line we used the fact that the distribution factorizes. 
\begin{figure}
    \centering
    \includegraphics[width=\linewidth]{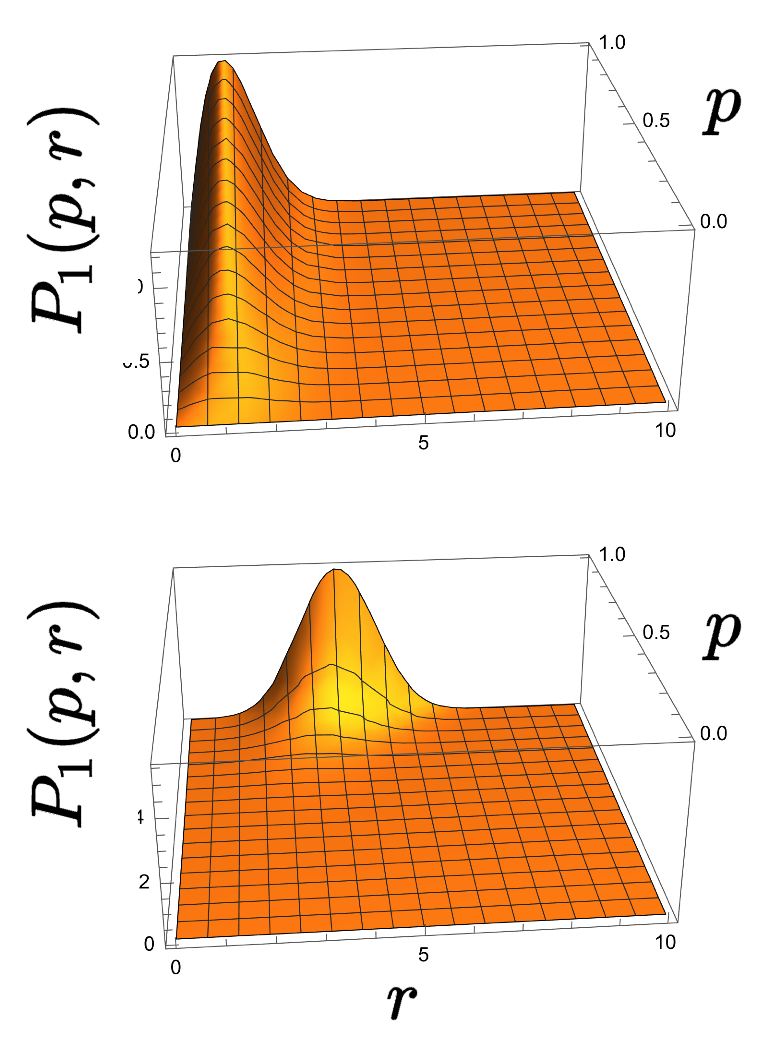}
    \caption{Distribution as a function of low $\kappa\approx 0$ and for $\kappa\approx 3.5$. We can see that the value of $r$ deviates from the low distribution. }
    \label{fig:distv}
\end{figure}

We then have, calling $\mathcal C( R)$ the enclosing disk with its boundary of radius $R$, we have
\begin{widetext}
\begin{eqnarray}
    P_1(p,r)&=&\int_{\mathcal C(R)} d\bm{x} \int_{\mathbb R^2}d{\bm{p}} \frac{e^{-\frac{\beta_T {\bm p_i}^2}{2 m}-\beta_L {\bm p}_i\wedge {\bm x}_i }}{\mathcal N_i} \delta(\sqrt{\frac{\beta_T}{m}}p-\|\bm p\|)\delta( R r-\|\bm x\|)\\
    &=&\int_0^{R} \tilde r d\tilde r \int_0^{2\pi} d\phi \int_0^\infty d\tilde p\ \tilde p \int_{0}^{2\pi} d\phi\ \frac{e^{-\frac{\beta_T {p}^2}{2 m}-\beta_L \tilde p \tilde r \sin(\theta-\phi)}}{\mathcal N_1} \delta(\sqrt{\frac{\beta_T}{m}}p-\tilde p)\delta( R r-\tilde r)
\end{eqnarray}
\end{widetext}
we now perform the change of variables $a=\theta-\phi$ with $-2\pi<a<2\pi$ and $b=\theta+\phi$, with $0<b<4\pi$ in the integral, with a jacobian satisfying $|\text{det}(J)|=2$, and $\tilde r=R  s$:
\begin{widetext}
\begin{eqnarray}
    P_1(p,r)
    &=&\int_0^{ R} \tilde r d\tilde r \int_0^{2\pi} d\phi \int_0^\infty d\tilde p\ \tilde p \int_{0}^{2\pi} d\phi\ \frac{e^{-\frac{\beta_T {\tilde p}^2}{2 m}-\beta_L \tilde p \tilde r \sin(\theta-\phi)}}{\mathcal N_1} \delta(\sqrt{\frac{\beta_T}{m}}p-\tilde p)\delta(Rr-\tilde r)\\
    &=&R^2 4\pi\int_0^{1} \tilde s d\tilde s  \int_0^\infty d\tilde p\ \tilde p \int_{-2\pi}^{2\pi} da\ \frac{e^{-\frac{\beta_T {\tilde p}^2}{2 m}-\beta_L \tilde p  R s \sin(a)}}{\mathcal N_1} \delta(\sqrt{\frac{\beta_T}{m}}p-\tilde p)\delta(Rr-R s)\\
     &=&R^2 4\pi \frac{r}{ R}  p \int_{-2\pi}^{2\pi} da\ \frac{e^{-\frac{{p}^2}{2 }-\beta_L R \sqrt{\frac{m}{\beta_T}} p  r \sin(a)}}{\mathcal N_1} \\
     &=& R \sqrt{\frac{\beta_T}{m}}  8\pi\ p e^{-\frac{{p}^2}{2 } } \int_{0}^{2\pi} da\ \frac{e^{-\kappa p  r \sin(a)}}{\mathcal N_1} =   \frac{ 16\pi^2   R}{\mathcal N_1}\cdot p r\ e^{-\frac{ {p}^2}{2 } } I_0(\kappa p r  ) 
\end{eqnarray}
\end{widetext}
where we defined $\kappa=\frac{\beta_L R \sqrt{m}}{\sqrt{\beta_T}}$, and used $\int_0^{2\pi} e^{-A \sin(x)} dx=2 \pi I_0(A)$, with $I_0(x)$ the Bessel function of the first kind.
We can calculate the normalization factor, as 
\begin{eqnarray}
    1=\frac{1}{ \mathcal N} \int_0^{\infty}dp\ p \int_0^1 dr\ r \int_0^{2\pi} d\theta e^{-\frac{ p^2}{2}-\kappa p r \sin(\theta)}\nonumber \\
\end{eqnarray}
with $\kappa=\frac{\beta_L R \sqrt{m}}{\sqrt{\beta_T}}$, and obtain
\begin{eqnarray}
     \mathcal N=\frac{2}{\pi}\frac{e^{\frac{\kappa^2}{2}}-1}{\kappa^2}
\end{eqnarray}
We can perform this integral exactly by expanding in powers of $\kappa$. We obtain the final result
\begin{eqnarray}
    P_1(p,r)=\frac{\kappa^2}{e^{\frac{\kappa^2}{2}}-1}\ p\ r\ e^{-\frac{p^2}{2} }I_0(\kappa p r)
\end{eqnarray}
with $p\in[0,\infty]$ and $r\in [0,1]$.

The distribution $P_1(p,r)$ is plotted in Fig. \ref{fig:distv} for small and larger values of $\kappa$.

\section{Order parameter and conserved quantities}\label{sec:chiparams}
Now, note that by construction, we have
\begin{eqnarray}
    L&=&\int dp~dr~d\alpha\ l(p,r,\alpha)P_1(p,r,\alpha)\\
    E&=&\int dp~dr~d\alpha\ e(p)P_1(p,r,\alpha)
\end{eqnarray}
where $l(p,r,\alpha)=p r \sin(\alpha)$ and $e(p,r)=p^2/2$. We can also obtain these by taking the derivatives with respect to $\beta_T$ and $\beta_L$. In particular, we can obtain these averages directly from the distribution, defining $\gamma=\kappa^2/2$ 
\begin{eqnarray}
    \langle  E \rangle&=&\langle \frac{u^2}{2}\rangle=\left(\frac{1}{e^{\gamma }-1}+1\right) \gamma\\
\langle  L \rangle&=&\langle p r \sin \alpha\rangle=-\frac{\sqrt{2} \left(e^{\gamma } (\gamma -1)+1\right)}{\left(e^{\gamma }-1\right) \sqrt{\gamma }}
\end{eqnarray}
In physical units, we have instead
\begin{eqnarray}
Z_1&=& \frac{2 \pi  \left(e^{\frac{m R^2 \beta _L^2}{2 \beta _T}}-1\right)}{\beta _L^2}\\
    \langle  E \rangle&=&-\partial_{\beta_T}\log Z_1=\frac{m R^2 \beta _L^2}{2 \beta _T^2 \left(1-e^{-\frac{m R^2 \beta _L^2}{2 \beta _T}}\right)}\nonumber \\
    \langle L\rangle&=&-\partial_{\beta_L}\log Z_1= \frac{2}{\beta _L}-\frac{m R^2 \beta _L \left(\coth \left(\frac{m R^2 \beta _L^2}{4 \beta _T}\right)+1\right)}{2 \beta _T}\nonumber 
\end{eqnarray}

Thus, we can define $\Xi(\gamma)$ as 
\begin{eqnarray}
    \Xi=\frac{\left(e^{\gamma } (\gamma -1)+1\right)^2}{\left(e^{\gamma }-1\right)^2 \left(\frac{1}{e^{\gamma
   }-1}+1\right) \gamma ^2}
\end{eqnarray}
where we used the fact that $m_i=m$.

\begin{figure}
    \centering
    \includegraphics[width=\linewidth]{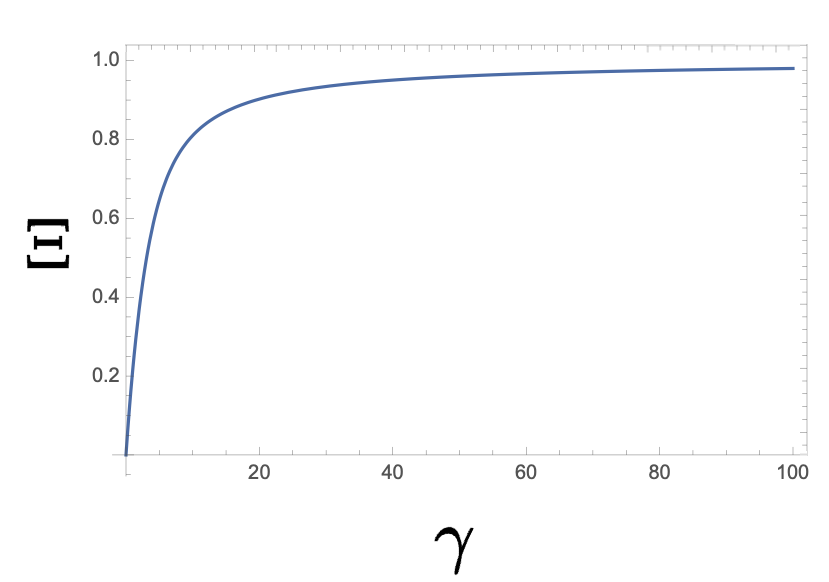}
    \caption{Function $\Xi$ as a function of $\gamma=\kappa^2/2$. For large values of $\gamma$, the monotonic function $\Xi$ converges to the value $1$.}
    \label{fig:plotchi}
\end{figure}

As a function of $\gamma$, the plot of $\Xi$ is shown in Fig. \ref{fig:plotchi}. Since the larger the value of $\kappa$, the larger the deviation from the Gibbs distribution, we can tell that, since $\Xi$ is a monotonic function, we can calculate the inverse $\kappa=\Xi^{-1}(x).$ 
For large values of $\gamma$, $\Xi\approx 1-\frac{2}{\gamma}$, which gives the approximate inverse near $\Xi=1$
\begin{eqnarray}
    \gamma\approx \frac{2}{1-\Xi}
\end{eqnarray}
and thus
\begin{eqnarray}
    \kappa\approx  \frac{2}{\sqrt{1-\Xi}}.
\end{eqnarray}
This implies that, when $\Xi$ approaches the values $1$, the distribution diverges substantially from Gibbs. 
Let us now discuss the marginal distributions over $p$ and $r$ only.
We have the marginal distributions
\begin{eqnarray}
    P_1^\gamma(r)&=&\int_0^\infty dp P_1(p,r)=\frac{2 \gamma  r e^{\gamma  r^2}}{e^{\gamma }-1}\\
    P_1^\gamma(p)&=&\int_0^\infty dp P_1(p,r)= \frac{\gamma  e^{-\frac{p^2}{2}} p \, _0\tilde{F}_1\left(2;\frac{p^2 \gamma }{2}\right)}{e^{\gamma }-1}
\end{eqnarray}
where $F_1(2;x)$ is  the regularized confluent hypergeometric function.

In the limit $\gamma\rightarrow \infty$ and $\gamma\rightarrow 0$, we have
\begin{eqnarray}
    \lim_{\gamma \rightarrow 0} P_1(r)&=& 2 r \label{eq:eq1}\\
    \lim_{\gamma \rightarrow \infty} P_1^\gamma(r)&=&\lim_{\gamma \rightarrow \infty}  2 r\gamma e^{-(1-r^2) \gamma} =2 \delta(1-r)\label{eq:eq2}
\end{eqnarray}
on the interval $[0,1]$. The factor of $2$ might be confusing but it is consistent. For eqn. (\ref{eq:eq1}), we have
\begin{eqnarray}
    \int_0^1 dr~ 2r=2\frac{1}{2}r^2|^1_0=1.
\end{eqnarray}
For eqn. (\ref{eq:eq2}) one needs to use
\begin{eqnarray}
    \int_0^1 \delta(1-r)dr=\frac{1}{2}
\end{eqnarray}
from the fact that $\delta$-function is located at the boundary of the integral,
\begin{eqnarray}
    \int_0^1 dr~ 2\delta(1-rr)=2\frac{1}{2}=1.
\end{eqnarray}

Thus, for $\Xi\rightarrow 1$ we have that our radial distribution is peaked on the boundary of the circle at equilibrium. On the other hand, we have
\begin{eqnarray}
    \lim_{\gamma\rightarrow 0} P_1^\gamma(p)=e^{-\frac{p^2}{2}} p
\end{eqnarray}
which is the Gibbs distribution. At the leading order for $\gamma$ large, we get instead
\begin{widetext} 
\begin{eqnarray}
   P_1^{\gamma \gg 1}(p)= \frac{\gamma  e^{-\frac{p^2}{2}} p \left(3 \sin \left(\sqrt{2} \sqrt{\gamma  \left(-p^2\right)}+\frac{\pi }{4}\right)-8 \sqrt{2} \sqrt{\gamma 
   \left(-p^2\right)} \cos \left(\sqrt{2} \sqrt{\gamma  \left(-p^2\right)}+\frac{\pi }{4}\right)\right)}{4\ 2^{3/4} \sqrt{\pi } \left(e^{\gamma
   }-1\right) \left(\gamma  \left(-p^2\right)\right)^{5/4}}
\end{eqnarray}
\end{widetext}
which is a distribution highly peaked at very large momenta.

\section{Further numerical simulations evidence}\label{eq:consmom}

\begin{figure*}
    \centering
    \includegraphics[width=\linewidth]{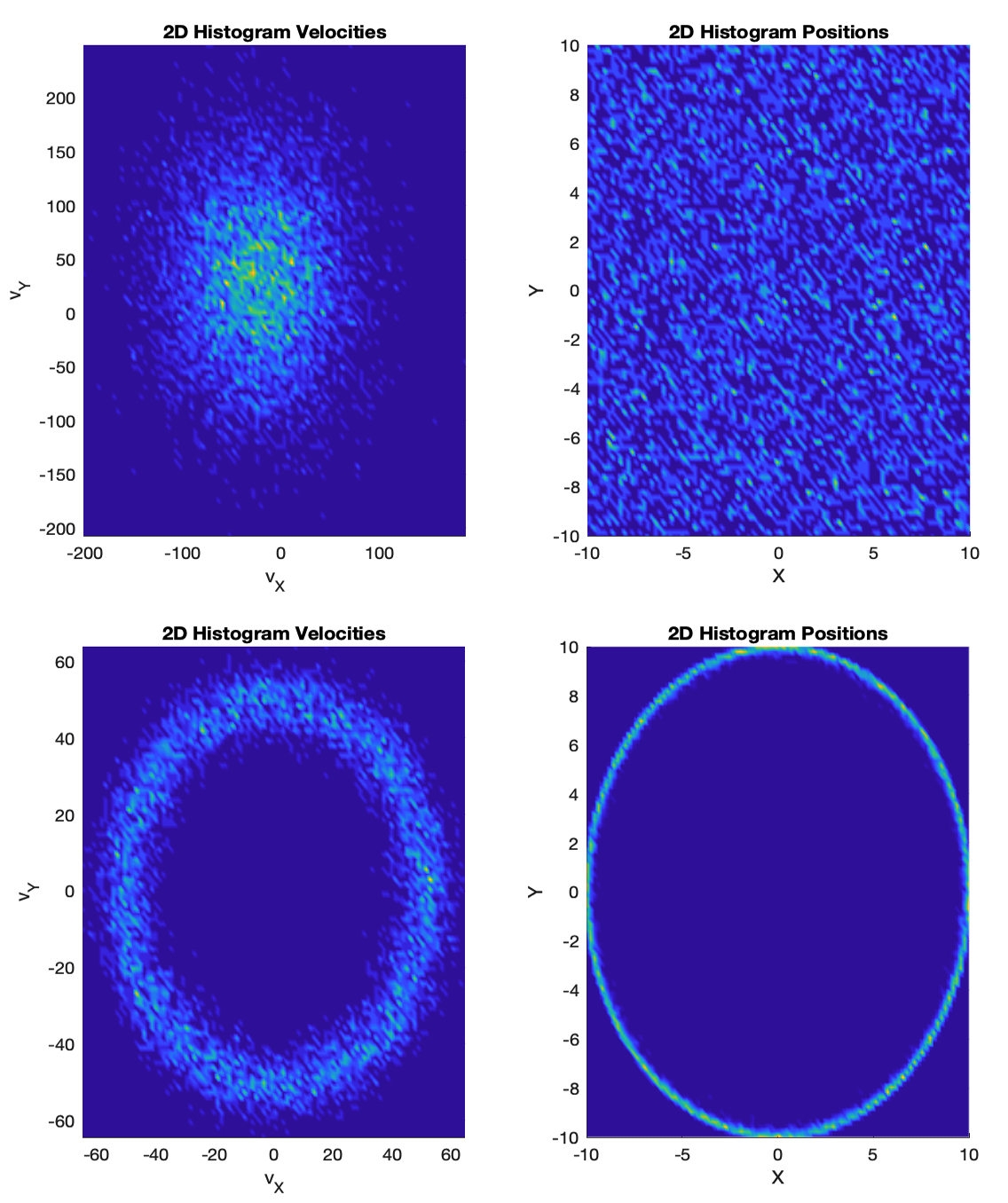}
    \caption{Results obtained with a time-driven simulation. \textit{Top}: Empirical 1-body final distribution of the velocity (left) and position (right) with 6000 disks for $\Xi_0=0.98$ with square boundary conditions. These distributions are consistent with Maxwell-Boltzmann, with radially distributed speeds, and uniform over the entire space of the system.
    \textit{Bottom}: Empirical 1-body final distribution of the velocity (left) and position (right) with 6000 disks for $\Xi_0=0.98$ with circular boundary conditions. These distributions deviate from Maxwell-Boltzmann. The velocity distribution is approximately zero for ${\bm v}=0$, while the radial distribution of particles is peaked around the boundary, exhibiting condensation. 
    }
    \label{fig:empdist}
\end{figure*}

\subsection{Details of the cluster simulations for TDMD} \label{ref:clust}

The codes we wrote initialize the positions and velocities of hard disks within the boundary according to an initial condition with an approximately fixed order parameter $\Xi$. 

\textbf{Choice of the initial conditions.} The initial conditions for this systems were the same as those described in the bulk of the paper.

\textbf{Time-driven simulation}. The time-driven simulation code, written in C, models a system of hard disks within a square and circular boundary. Time is divided into small increments $\delta t$. At each step:
a) We update the positions of all particles based on their velocities b) We check for overlaps (collisions) between particles c) We resolve collisions by adjusting velocities and rolling back positions.
This method, though straightforward, is computationally intensive due to frequent collision checks and fine time increments. In the case of hard disks, in addition, the criticism is that an addition time parameter is introduced in the model the time step $dt$, and that is not completely clear that one is truly simulating hard disks. However, in the presence of long-range interactions between particles, EDMD cannot be used. We use this other simulation technique, again, to stress that the effect is not dependent on the particle's interaction.  

\begin{figure}
    \centering
    \includegraphics[width=\linewidth]{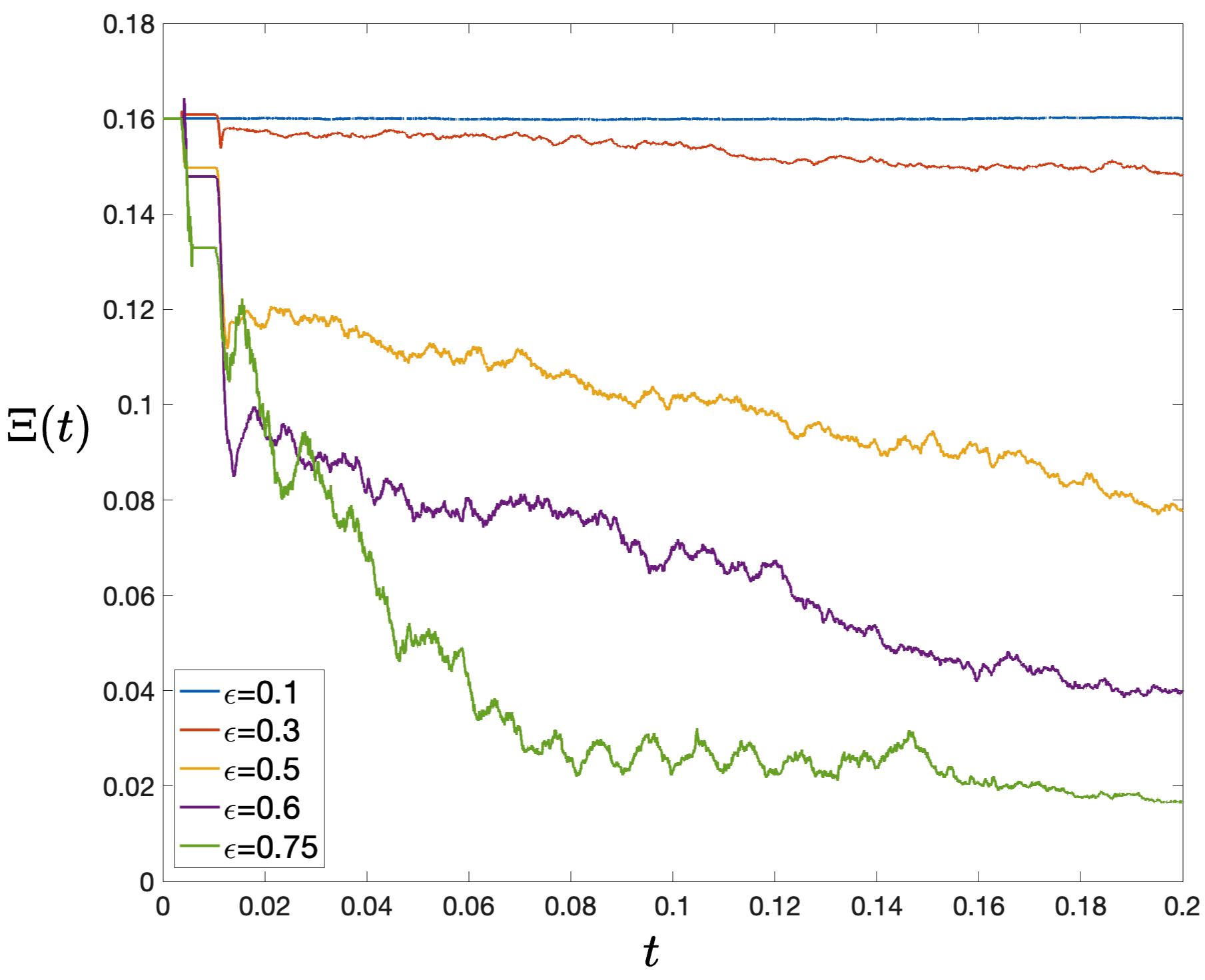}
    \caption{The order parameter $\Xi(t)$ as a function of time and the eccentricity. For zero eccentricity of the boundary (circle), we see that it is conserved, but as soon as the eccentricity is non zero, at each collision with the boundary a small amount of angular momentum is lost. }
    \label{fig:eccchit}
\end{figure}

To analyze the long-time distribution of the system of hard disks, we integrated numerically the time evolution of each disk.  Regarding the boundaries of our simulations, we considered square, circle, and elliptical boundaries.  For the case of the circle and square boundaries, the numerical results on the empirical one-body distributions can be found in  Fig. \ref{fig:empdist} (top) in the long-time regime. We have considered the radius of the boundary $R=10$ (or linear size in the case of the square), and the radius of each disk to be $r\propto \frac{\sqrt{N}}{R}$, so that $\eta=A_{particles}/A_{b}=\frac{\pi N r_d^2}{A_b}=0.1$ is a constant well below the critical density $\eta\approx 0.70$.  We have used the square boundary conditions to calibrate our code. Unsurprisingly, for the case of square boundaries, the distribution is well-approximated by a Maxwell-Boltzmann distribution in momenta with the peak of the angular momentum located at $p=0$, and flat in the coordinates. We found consistent results for the initial distribution with high and low total angular momentum.
We have found Maxwell-Boltzmann distributions in all ranges of $\Xi$ for the case of the square boundary, as expected. For the case of the circle, instead, numerical simulations show in Fig. \ref{fig:empdist} (bottom) that at equilibrium the one body distributions, in particular when $\Xi\approx 1$, strongly deviate from the Maxwell-Boltzmann. In particular, we find that for $\Xi\approx 1$ the asymptotic one-body momentum distribution $P_1({\bm p})$ is zero for ${\bm p}\approx 0$. Similarly, the one-body coordinate distribution $P_1({\bm r})$ stops being flat and indeed is highly concentrated near the boundary. Both distributions respect the rotational symmetry of the boundary. 
In order to test whether the parameters of our simulations were sufficient to maintain angular momentum, in App. \ref{sec:decayop} we plot the order parameter $\Xi(t)$ for both the circle and the square, showing that this is constant for the circle and it decays exponentially for the case of the square. Since the notion of thermalization is associated with the asymptotic emergence of the Gibbs distribution, or alternatively Maxwell-Boltzmann, we refer to this scenario as a breakdown of thermalization as a result of the conservation of angular momentum. Intuitively, one can see that this order parameter must be valid only in the liquid phase we are focusing on, as it is independent of the packing fraction $\eta$.
More specifically, there are two main events that we consider. First, we identify when disks overlap with the boundary, and perform a perfect reflection (see Fig. \ref{fig:collisions} left). The total energy $E_i=\frac{\|{\bm p}_i\|^2}{2m}$, where ${\bm p}_i$ is the momentum of the particles, but momentum is not conserved if not in magnitude. We also identify when particles collide, using the infinite potential between the particles. As a result, we identify when the disks overlap, and update their momenta, satisfying $E_{pre}=E_{post}$ and ${\bm p}_{pre}={\bm p}_{post}$, e.g. a fully elastic collision.  Disks then follow straight lines in between collisions. 
 In all our simulations, we have varied the time step until we observed an angular momentum constant over the whole simulation, and ultimately have chosen a time step $dt=10^{-6}$, with an effective total time of $T=3000$. We performed the numerical simulations with $N=6000$ disks, performed on a cluster \footnote{We stress that the relatively small number of disks is because we are not performing a Monte Carlo sampling but a dynamical simulation of the disks' collisions.}. This means that we allow for a small overlap of the disks, but momentum and energy are conserved in disk collisions. 
The code was parallelized using MPI directives for multi-core utilization on the LANL Darwin cluster, distributing disks among threads for concurrent Monte Carlo steps. The simulation was submitted as a batch job using Slurm for high-performance computing, utilizing multiple nodes or cores. The results can be summarized in the two figures of Fig. \ref{fig:empdist}. As we can see, the 1-body distributions exhibit the condensation phenomenon as in the case of the time-driven simulation.
It is interesting to see what happens when deviations from the perfect circle occur. In this case, at least one obtains a form of weak ergodicity breaking, in which the configuration is stuck for long times in a microcanonical angular momentum state. The simplest example is the case of an ellipse. If $R_{min}$ and $R_{max}$ are the lengths of the minor and major axis of the ellipse, a parameter that controls the degree to which the angular momentum is conserved is the eccentricity, defined as $\epsilon=\sqrt{1-R_{min}^2/R_{max}^2}$. For $\epsilon=0$ we are in the situation of a perfect circular boundary. In Fig. \ref{fig:eccchit} we plot the parameter $\Xi(t)$ for various values of the eccentricity. As it can be seen, for $\epsilon=0.1$ the order parameter remains constant for up to $T=2\cdot 10^4$ numerical steps. For larger values of $\epsilon$, the decay is noticeable.


\subsection{Boundary collisions EDMD} \label{sec:edmd}


Compared to other standard event-driven simulations, the only difference is how we calculate the collision time for a circular boundary. For a square boundary, the time of collision with the boundary only depends on the $x$ and $y$ components of the velocity, and thus it is simply $t_c=min(x_0/v_x,y_0/v_y)$. For a circular boundary, one needs to search for an intersection. For a particle located inside the circular boundary, in ${\bm r}_0=(x_0,y_0)$, and with speed $(v_x,v_y)$. 
Consider a particle of mass $m$ moving with velocity ${\bm v} = (v_x, v_y)$ towards the circular boundary of radius $R$ centered at the origin. We want to find the time $t$ until the particle collides with the boundary. The reflection from the boundary then occurs exactly using the formulae from the previous section.

The equation of motion for the particle is given by the straight line trajectory:
$$
{\bm s}(t) = \mathbf{r}_0 + \mathbf{v} t.
$$

The distance of the particle from the origin at time $t$ is:
$$
s(t) = \sqrt{(x_0 + v_x t)^2 + (y_0 + v_y t)^2}
$$

The particle collides with the boundary when $r(t) = R-r$. Substituting $s(t) = (R-r)$ into the equation above and squaring both sides, we get:
$$
(x_0 + v_x t)^2 + (y_0 + v_y t)^2 = (R-r)^2
$$

Expanding and rearranging terms, we obtain the quadratic equation:
$$
(v_x^2 + v_y^2) t^2 + 2(x_0 v_x + y_0 v_y) t + (x_0^2 + y_0^2 - (R-r)^2) = 0
$$

This is a quadratic equation in $t$, which can be solved using the quadratic formula:
$$
t = \frac{-b \pm \sqrt{b^2 - 4ac}}{2a}
$$
where $a = v_x^2 + v_y^2$, $b = 2(x_0 v_x + y_0 v_y)$, and $c = x_0^2 + y_0^2 - (R-r)^2$.

We take the positive root (since we are interested in the time until the next collision). Thus, the collision time $t_v$ between the particle and the boundary is given by:
\begin{eqnarray}
t_v 
&=&- \frac{{\bm r}_0\cdot {\bm v}}{\|{\bm v}\|^2}+\sqrt{\frac{({\bm r}_0\cdot {\bm v})^2}{\|{\bm v}\|^4}-\|{\bm r}_0\|^2-(R-r)^2}
\end{eqnarray}

This equation gives us the time until the particle collides with the circular boundary.

\begin{figure}
    \centering
    \includegraphics[width=\linewidth]{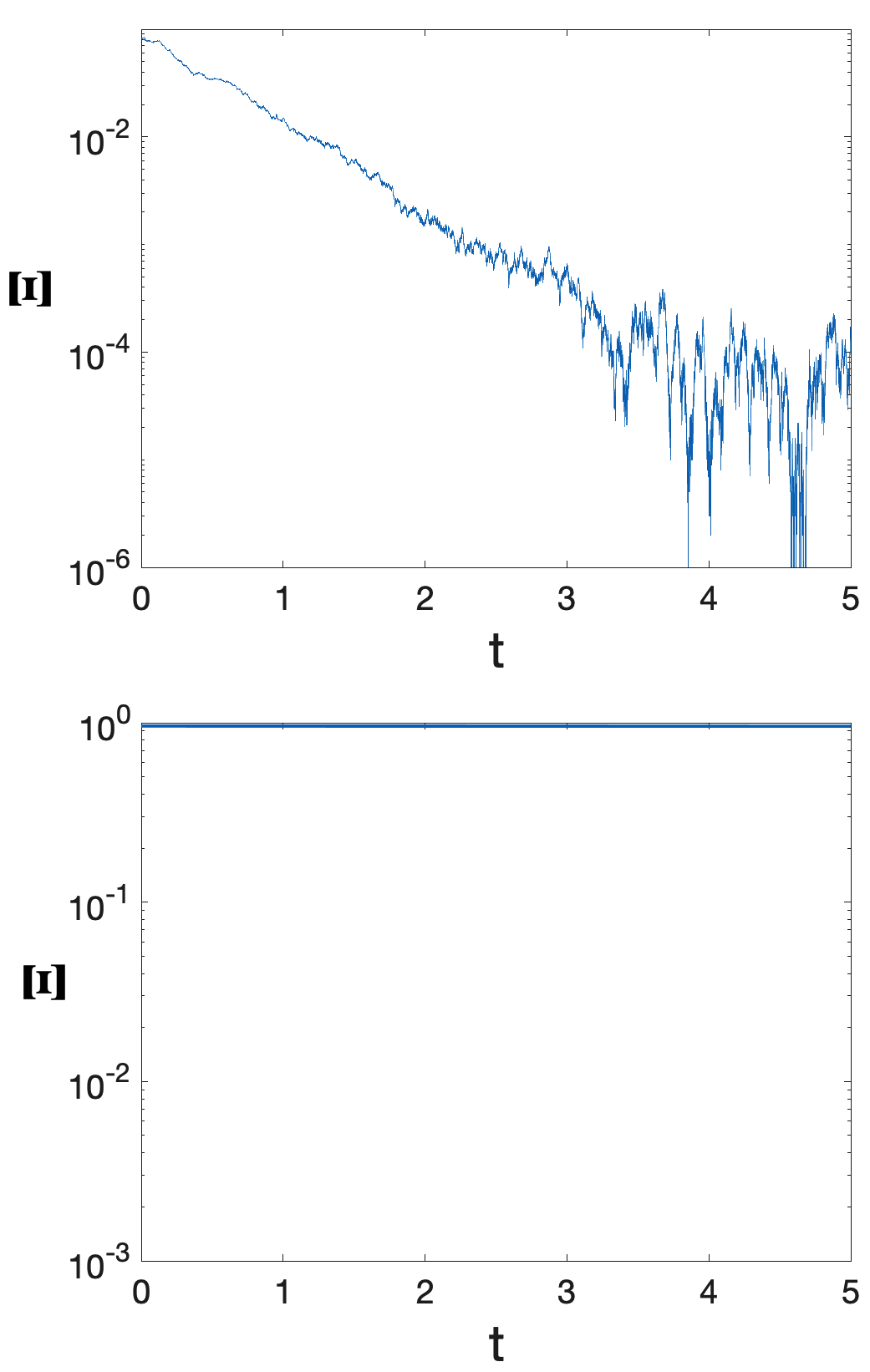}\\
    \caption{Top: Decay of the parameter $\Xi$ for a square boundary. Bottom: the constancy of the parameter $\Xi$ for t for the simulations on the circle.}
    \label{fig:LogChi}
\end{figure}

\section{Decay of the order parameter}\label{sec:decayop}
The order parameter $\Xi$ is constant if both energy and angular momentum are conserved. If the angular momentum is not conserved, then we expect $\Xi\rightarrow 0$ as a function of time. Since in particle collisions angular momentum is conserved, the loss of angular momentum is due to the reflection from a boundary.
We note that
\begin{eqnarray}
    \frac{d}{dt} \Xi(t)=\Xi(t) 2 \frac{d}{dt} \log L(t)
\end{eqnarray}
Now, during an infinitesimal time, we have that 
\begin{eqnarray}
    \frac{d\log L}{dt}=-\mathcal L \nu dt
\end{eqnarray}
where $\mathcal L(t)$ is the loss of angular momentum per collision, while $\nu$ is the expected number of collisions in the unit of time $dt$ between particles and the boundary.

As a result, we have
\begin{eqnarray}
    \frac{d\log \Xi}{dt}=-2 \mathcal L \nu 
\end{eqnarray}
and we obtain that
\begin{eqnarray}
    \Xi(t)=\Xi_0 e^{-2 \mathcal L \nu t}
\end{eqnarray}
We can see, for a square boundary, how the order parameter decays as a function of time in Fig. \ref{fig:LogChi}.


\end{document}